# Unconventional Logic Elements on the Base of Topologically Modulated Signals

Guennadi A. Kouzaev, Igor V. Nazarov, Andrew V. Kalita

Laser and Microwave Information Systems Dept., Moscow State Institute of Electronics and

Mathematics

3/12 Bol. Trekhsvaytitelsky per. Moscow,110028 Russia

e-mail: kouzaev@mail.ru, g132uenf@hotmail.com, www.topolog.da.ru

## ABSTRACT

The paper presents new results in the field of super high-speed and multi-valued signal processing Writting digital information into spatial structures (topological charts) of electromagnetic field pulses. allows to use passive circuits for fulfillment several subpicosecond spatial logical operations. This is confirmed by analysis of several physical effects in solids and micron circuits, which influence on time delay of signals. A subpicosecond circuit for spatially modulated signal switching is considered.

An analogy between electromagnetic mode physics and several aspects of quantum mechanics is studied. On this base a new digital multi-valued device for spatially modulated signal processing is suggested and modeled. A conclusion on possibility to design a new threedimensional architecture of super –density IC has been made.

# 1. INTRODUCTION

One from final stages of development of the electronic integrated circuits is the passage to three-dimensional quasicontinuum medium for processing electromagnetic signals having complicated spatially - temporarily forms. Indication of this tendency is the minimization of interelement distances in three-dimensional VLSI, diminution of sizes of the active elements and magnification of velocity of their work [1,2]. The signals in new super high-speed ICs become three-dimensional spatial objects which are capable to carry digital or analog information by the space structures of pulse fields. The tendency allows to use some analogies to optical space methods for processing the electromagnetic signals in electronic ICs [3-9].

The first results in this direction were obtained in outcome of study of space structures of fields in the microwave three-dimensional integrated circuits, which were suggested and designed for airspace engineering [10]. It was shown, that the structure of force lines of fields (topological chart) is capable to carry discrete information [3-5]. The spatial field characteristics are changed discretely during diffraction of modes or package of modes on passive discontinuities. The further researches have shown a possibility of design full series of the Boolean logic elements for microwave spatially - modulated signals [4,9], and micron sized subpicosecond passive components which are capable to switch digital signals to different layers of the integrated circuits or to work as binary matched space filters [11,12]. For design such circuits have appeared useful some analogies to optical methods of processing spatially modulated signals.

The purpose of the paper is study of multi-valued nature of the recently (1992) suggested electromagnetic signals (digital images) and modeling new devices on this base.

## 2. HIGH DENSITY INTEGRATED CIRCUITS AS QUASI-OPTICAL DEVICES

Let's define a likeness and difference of optical and electromagnetic signals for correct application of method of analogies to high-density IC [13]. The optical signal represents an impulse sinusoidal field by temporal duration up to units femtosecond. The electromagnetic signal in electronic circuits can look like sinusoidal segments of microwave oscillations or to represent impulse, which form is close to a rectangle. Its band of temporal frequencies can take a spectrum from 0 up to units terahertz.

 The band of space frequencies of an optical signal, as a rule, is much wider than similar performance of an electromagnetic signal. Space information can be transmitted by a ray in open space or by light waveguide. The signal in electronic circuits on strip transmission lines is capable to be spatially-modulated only under existence of multimode condition, for example at use of coupled transmission lines (Fig. 1). Thus the magnitude of the signal space spectrum is determined by number of propagation modes or number of strip conductors. Dimensions of separate components of the electronic circuits are significant less than the least wavelength of the signal and they are elements of "near field zone", where the quasistatic six-component electromagnetic fields prevails. The effects of superposition of electromagnetic fields with origin of a fractal space structure of potentials are characteristic of this space area of electromagnetic field. The signal processing in this area is produced by the discrete active and passive elements and a possibility of separate transformation of magnetic and electrical fields is realized.

Behind of nanocircuits, the optical components are devices using the wave mechanism of interference and diffraction in the far zone. It is possible to name the circuits as "far field" components. The active processing optical signals is possible only by using special nonlinear wave mediums or optoelectronic devices [14].

Thus, the common electromagnetic nature of signals in optic and electronic circuits allows to apply similar methods of their spatial processing. But the possibilities of electronic circuits in relation to optical engineering can be essentially large at the expense of significant variety of element basis [12-15-18].

# 3. SPATIALLY MODULATED SIGNALS IN ELECTRONIC IC AS TOPOLOGICAL OBJECTS

During development the electronic circuits of this type the fact of low dimensionality of space spectrums of electromagnetic signals in electronic circuits was taken into account. The best kind of the signal has been recognized a pulse with discrete modulation of its electromagnetic field structure [3,4,19]. Thus the information carrier is the topology or topological chart of a picture of force lines representing an ordered combination of basic elements of the picture: separatrixes, positions of field equilibrium. The topological chart is possible to name as some kind of a quantum of spatial information According to the nature of the topology, the characteristic is capable only to discrete modifications [3,10]. The theory of such signals and methods of their processing have been already considered in [3-6,8,9]. Its basic sense consists that the field topology can be changed discretely during diffraction or interference of electromagnetic waves. The effects, as well known, are linear operations concerning amplitudes of signals, but not topology of fields [8,11]. Therefore, the part of discrete logic operations can be fulfilled by passive integrated structures like in optical engineering. Small inertia of the passive circuits allows to process the vector images (topological charts) of electromagnetic fields practically in real time. The techniques of the signal processing is possible to name as topological computing.

# 4. PHYSICAL BASEMENT FOR SUPER HIGH SPEED SIGNAL PROCESSING BY PASSIVE ELECTRONIC COMPONENTS

The writing digital information to a space structure of electromagnetic fields allows to realize some high speed logic operations on the base of using diffraction effects and fast-response equilibrium processes in conductors having continuous energy spectrums [11,12]. These effects are divided into two basic groups (Table 1). The first from them are composed from phenomena, having a place in solid states. For realization subpicosecond operation it should pay special attention on time characteristic of using materials. The second group is stipulated by macroeffects, having electromagnetic nature. For example, the special attention should be given to transients distorting the forms of switched impulses. The method of equivalent circuits [22] which are taking into account parasitic reactivities of discontinuities of strip transmission lines, was applied to an evaluation of duration of these processes (Fig. 2). It has been shown, that parasitic reactivities of the elementar discontinuities are the cause of transients, which duration are in limits from 0.01 to 0.2 picosecond, if the sizes of strip transmission lines are about several microns.

Thus, the system analysis of main effects in strip transmission lines and circuits (Table 1) allows to make a conclusion about a possibility of realization subpicosecond operations with spatially - modulated signals by passive components, which have dimensions about several microns.

## 5. BINARY CIRCUITRY FOR TOPOLOGICAL SIGNAL PROCESSING

The basic concepts of switching spatially modulated signals (Fig.1), are constructed on the principle of matched space filtration of impulses, the structures of which fields are varied discretely [5].

On Fig. 3, a circuit of the resistive switch for spatially modulated field signals and its truth-table are represented. Input pulses of even or odd modes of coupled strip transmission lines are switched to different outputs due to using the mechanism of matched spatial filtration (Fig.3, b). The resistors in the circuit allow to ensure a short aperiodic condition of transients at switching the signals. Minimization of the difference of even and odd modes speeds is achived due to using symmetrical construction of the coupled strip transmission lines [5-9].

On Fig.4 the switched signals are represented. The duration of transients did not exceed a several tenth long of picosecond for rectangular entering signals. Essentially smaller distortions are appeared if signals of the Gaussian form are used, the duration of which front makes about 1 picosecond. On the functions the considered switch is equivalent to the current transistor switch containing several transistors. The comparative analysis (Fig. 5 ) of ohmic power losses at transistor gates and offered switch indicates an advantage last [5,6-9,23,24].

The inclusion in the circuits for space signals processing the active elements (Fig. 6) allows to achieve new functional advantages [8,25]. For example, the logic processing of amplitudes of impulses can be conjugate with discrete matched space filtration. The given type of evolution of the logic circuits is capable to design active, controlled analogs of hologram devices on the electronic basis [8,12,20].

The developed circuits were experimentally studied. The first results were obtained in microwave range for switches of different types. [4,9,23,24]. Another experimental results were for the digital signals and circuits with clock frequency about several megahertz [11,16,18,21,25,26]. On the base of scaling method a conclusion made on possibility to design a resistive switch for switching topologically modulated signals with clock frequency several hundred gigahertz. The last results are touched to developing and application picosecond generator of topologically modulated signals, designed by G. Domashenko [13].

## 6. MULTI-VALUED NATURE OF TOPOLOGICALLY MODULATED SIGNALS AND THEIR APPLICATION FOR MODELING QUANTUM LOGIC ELEMENTS

One of the most perspective methods of signal processing is application of the quantum algorithm and quantum devices. The approach uses physical parallelism for effective signal computing [27,28]. From an algorithmic point of view, the quantum logic circuits differ from the classical Boolean devices only by possibility of realization special type of multi-valued logic [29]. Besides in the quantum elements the superposition of quantum states corresponds to a new logic level. The last feature allows, for instance, to solve the well known problem of exponential complexity of calculations [29].

Comparison of quantum mechanics and electromagnetic mode physics in waveguides has been shown existence of an anology between them from the point of formal algorithmic view. For this purpose it is enough to compare multi-level energy diagram of a quantum element and multi-level frequency diagram for modes of a waveguide. Each mode has own topology of electromagnetic field. The mode topology is some kind of quantum of space information for mode and may correspond to a logical level. During superposition of modes, the topology (topological chart) of the modes may be changed abruptly, creating a new logical level of information system, like in a quantum element [8]. Besides, the topologically modulated signals carry digital information by their magnitudes. It allows to realize special types of multi-valued logic signal processing. The first results in this field were obtained in [4,5,7,8,30]. On the base of this physics a hybrid logic devices were developed and studied: NOT, OR/AND logical circuits ( the state of the last may be changed thanks to variation of amplitudes of comparing topologically modulated signals). On the base of the hybrid logical circuits a microwave trigger was suggested for processing the multi-valued signals of this type [4].

In [8,13,28,29, 31] a multi-valued logic circuit was suggested and modeled (Fig. 6,7). The device allows to process information contained in magnitudes and topological charts of the field signals due to diodes on the logical outputs of above considered passive switch (Fig.6, b ). The full number of logical states of topologically modulated signals in coupled strip transmission line is four . So the circuits may be pertinent for 4-value logical circuitry. (Fig.6, b). The results of modeling the multi-valued switch are shown on Fig.7. The switch transients depend, practically, only on parasitic reactivities of the used microwave diodes. It is obviously, that amplitudes of output signals depend on polarities and topological charts of input signals.

An advantage of the type of multi-valued signal is a possibility to realize the logical circuits without using an additional energy for supporting these logical levels opposite to well-known amplitude multi-valued techniques. So, the application analogies of mode physics to quantum mechanics allows to develop and design new type circuits having multi –logic and quasi-neural features. The circuitry opens a possibility of modeling quantum algorithm on the base well-known technology of IC producing.

## 7. A VIEW ON PERSPECTIVE HIGH - DENSITY INTEGRATED CIRCUIT ARCHITECTURES

The increase of density of IC dictates to use spatial methods of signal processing in electronics. On this way the most perspective approach is development new IC elements and  physical algorithms, allowing to achieve a new grade of density and new quality in signal processing. One of these approaches is a topological computing, dealing with discretely modulated  spatial signals. Due to that it is possible to develop super-high speed circuits, realize multi-valued and pseudoquantum devices and consider a future architecture of integrated electronic circuits as a threedimensional medium - new type of the artifical, (designed) electronic hologram [8,12,32,33].

## 8. CONCLUSIONS

The paper contains original and overviewed results in the field of new logical signals – topologically modulated images of electromagnetic fields, carrying information by their magnitudes and field structures. It has been shown multi-valued nature of the signals and existence of "electromagnetic field topological logic". Part operations from this logical system may be similar to the quantum logic. The suggested digital devices are able to process digitally modulated signals with subpicosecond time delay (passive components). It has been shown a formal possibility to simulate quantum logic operation by threedimensional circuits for topologically modulated signals. The combined circuits for parallel processing magnitude and topological information are suggested as a base for multi-valued signal devices.

The considered approach (topological computing) allows to develop new threedimensional integrated circuits as an information processing medium with new possibilities.

| ₁ | Physical effect | Time or frequency evaluation of an effect |
|---|---|---|
| 1. | Limited mode velocity in microstrip transmission lines. Time delay of signal in a microstrip transmission line on the substrate with dielectric permittivity $\varepsilon$: | ~ 33.3 $\sqrt{\varepsilon}$ , fs/ ***m***m |
| 2. | Inertia of interaction of electromagnetic field with free charge in the region of low values of photon energy. | Defined by efficient or free mass of charges |
| 3. | Maxwell relaxation time of charges in conductors: | ~ 0.001 – 0.01 fs |
| 4. | Collective effects in the electronic plasma. Period of plasma frequency in the conductors: | ~ 0,067 – 0,2 fs |
| 5. | Relaxation phenomenas in dielectric. Time constant of electronic polarization: Time constant of atomic polarization: | ~ 1 – 10 fs ~ 10 – 10000 fs |
| 6. | Minimal time of transition an electron from one energy level on the another in atom: | ~ 1 – 10 fs |
| 7. | Typical theoretical time of electron relaxation in quantum nanoelements: | 100-1000 fs |
| 8. | Electron-phonone interaction. Resonant frequency in conductors: | ~ 10 THz |
| 9. | Transient-time effects on discontinuities of strip transmission lines of micron sizes. Typical duration time of transient process on discontinuities (Idealized Oliner model for discontinuities): | ~ 80-150 fs |
| 10. | Excitation of higher modes on discontinuities of microstrip transmission lines in VLSI. Cut-off frequency of the first higher mode: | ~ 10- 100 THz |
| 11. | Excitation of surface waves in micron microstrip transmission lines. Critical coupling frequency of the strip and surface modes: | 10-100 THz |
| 12. | Limited mode velocity in microstrip transmission lines. Time delay of signal in a microstrip transmission line on the substrate with dielectric permittivity $\varepsilon$: | ~ 33.3 $\sqrt{\varepsilon}$ , fs/ ***m***m |
| 13. | Inertia of interaction of electromagnetic field with free charge in the region of low values of photon energy. | Defined by efficient or free mass of charges |
| 14. | Maxwell relaxation time of charges in conductors: | ~ 0.001 – 0.01 fs |
| 15. | Limited mode velocity in microstrip transmission lines. Time delay of signal in a microstrip transmission line on the substrate with dielectric permittivity $\varepsilon$: | ~ 33.3 $\sqrt{\varepsilon}$ , fs/ ***m***m |
| 16. | Inertia of interaction of electromagnetic field with free charge in the region of low values of photon energy. | Defined by efficient or free mass of charges |
| 17. | Maxwell relaxation time of charges in conductors: | ~ 0.001 – 0.01 fs |
| 18. | Collective effects in the electronic plasma. Period of plasma frequency in the conductors: | ~ 0,067 – 0,2 fs |
| 19. | Relaxation phenomenas in dielectric. Time constant of electronic polarization: Time constant of atomic polarization: | ~ 1 – 10 fs ~ 10 – 10000 fs |
| 20. | Minimal time of transition an electron from one energy level on the another in atom: | ~ 1 – 10 fs |
| 21. | Typical theoretical time of electron relaxation in quantum nanoelements: | 100-1000 fs |

Figure legends

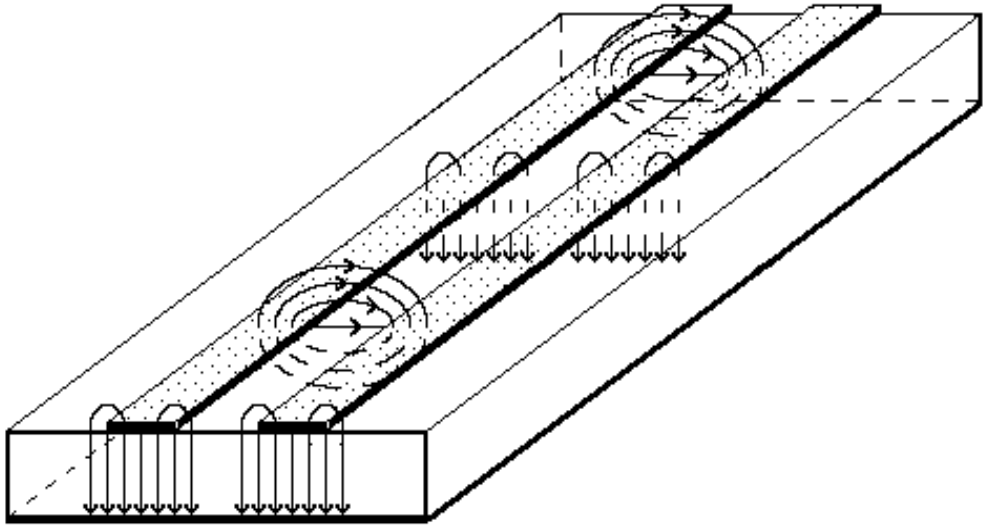

Fig. 1. An example of topologically modulated field pulses in coupled microstrip transmission lines. Digital information may contain in pulse field structures -topological charts (two logic levels) and in their amplitudes (other two logical levels).

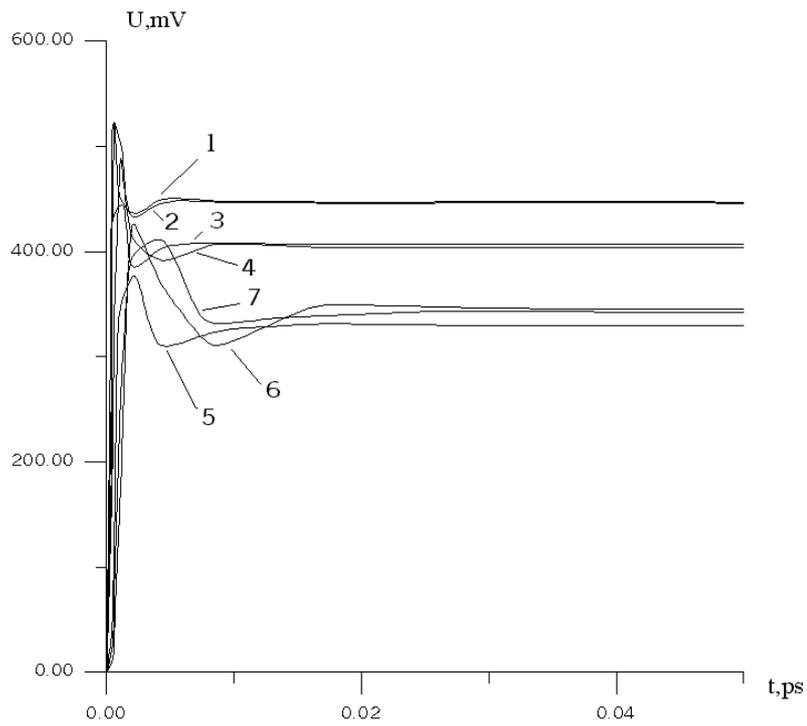

Fig.2. Transients on steps of microstrip transmission lines. ε-dielectric permittivity of substrate, h=1.3 mkm - thikness of substrate, $W_{1,2}$ - width of strip conductors, T- duration period of transients.

| Curve number | $W_1$ | $W_2$ | ε | T |
|---|---|---|---|---|
| 1 | 1.0 | 1.5 | 3.5 | 0.0011 |
| 2 | 1.0 | 1.5 | 9.6 | 0.0021 |
| 3 | 0.5 | 1.0 | 3.5 | 0.0035 |
| 4 | 0.5 | 1.0 | 9.6 | 0.0047 |
| 5 | 0.3 | 1.0 | 3.5 | 0.0077 |
| 6 | 0.5 | 1.5 | 9.6 | 0.0118 |
| 7 | 0.5 | 1.5 | 3.5 | 0.0132 |

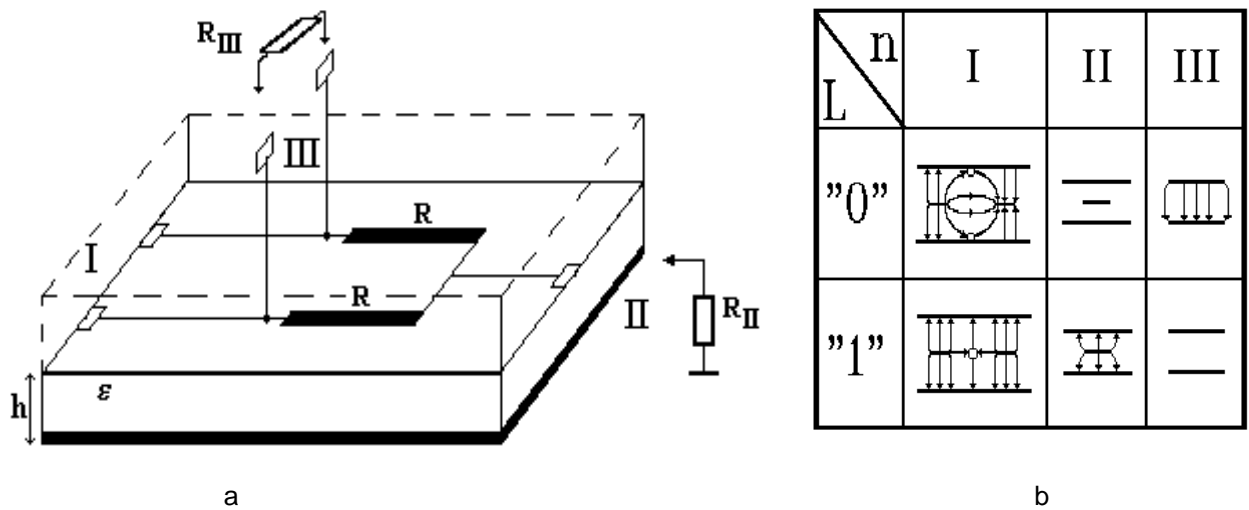

a                                                              b

Fig. 3. Binary switch for topologically modulated pulse field signals (a) and its truth-table (b):

I - input of the signals (coupled strip transmission lines with characteristic impedance $Z_e$ and $Z_o$),

II - output of logical "1" ( a two conductor transmission line with characteristic impedance $R_{II}$ ),

III - output of logical "0" (a strip transmission line with characteristic impedance $R_{III}$.

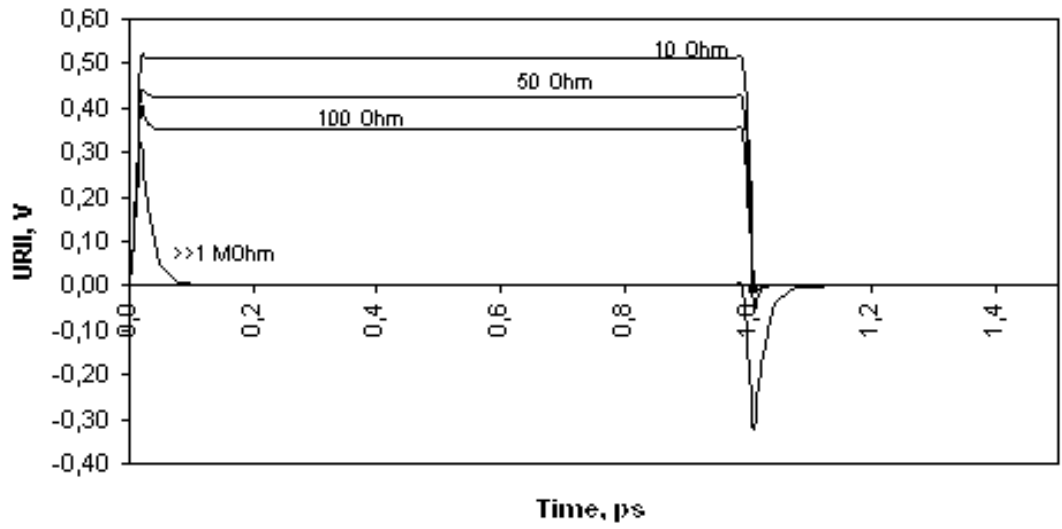

a

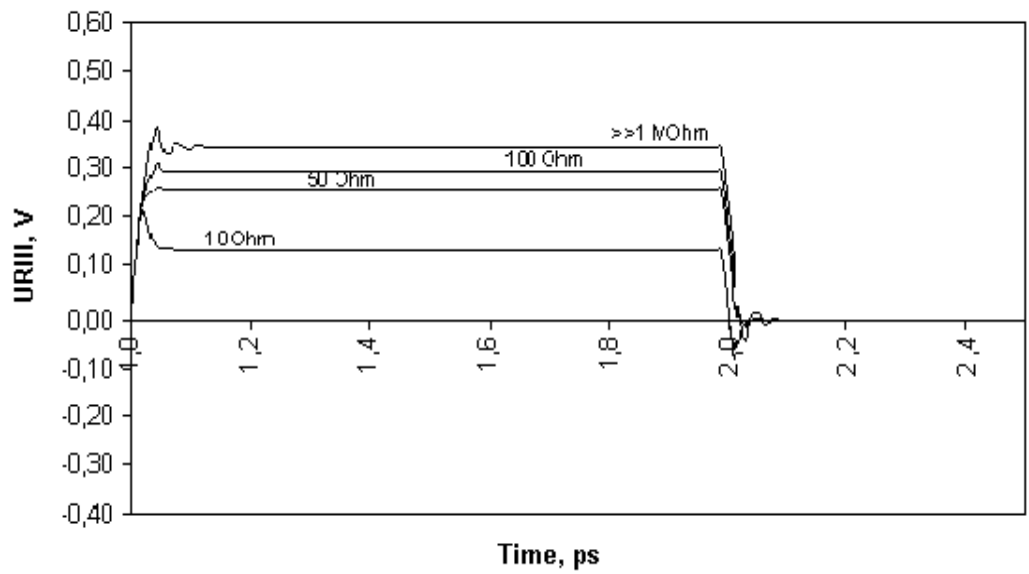

b

Fig. 4. Impulses of odd (a) and even (b) modes, transmitted the switch. R is parametrically varying constant. $R_{ii} = R_{ii} = 50$ Omh, $\varepsilon = 3.5$, h=3 mkm, $\Delta l$=1 mkm, w=1 mkm, s=1 mkm.

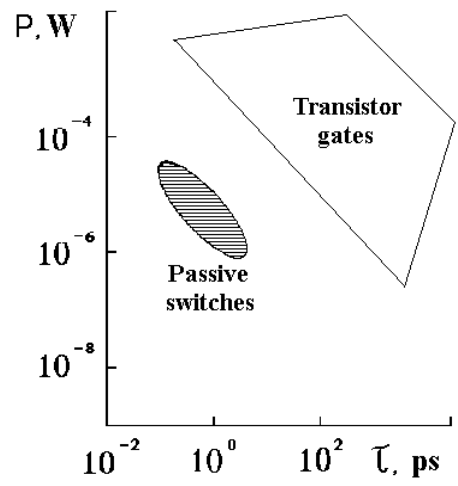

Fig. 5.Time-energy performances of passive switches for topologically modulated field signals and their transistors analogs.

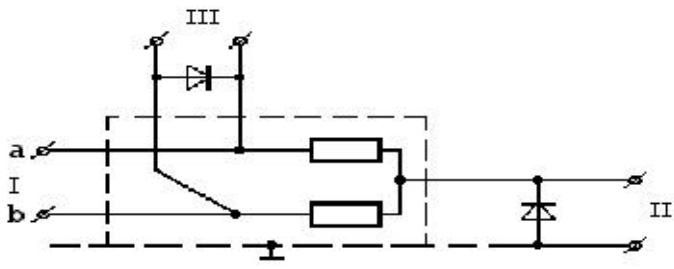

| Input signal | | Output signals | |
|---|---|---|---|
| I | | II | III |
| Even mode | $U^a=U^b<0$ | 0 | 0 |
| | $U^a=U^b>0$ | 1 | 0 |
| Odd mode | $U^a>U^b$ | 0 | 0 |
| | $U^a<U^b$ | 0 | 1 |

a                                        b

Fig. 6. A switch for multi-valued signal processing (a). Signal information contains in structure and amplitudes of field pulses. I –input coupled strip transmission lines, II- output strip transmission lines, III-two conductor transmission. Truth table for the switch (b).

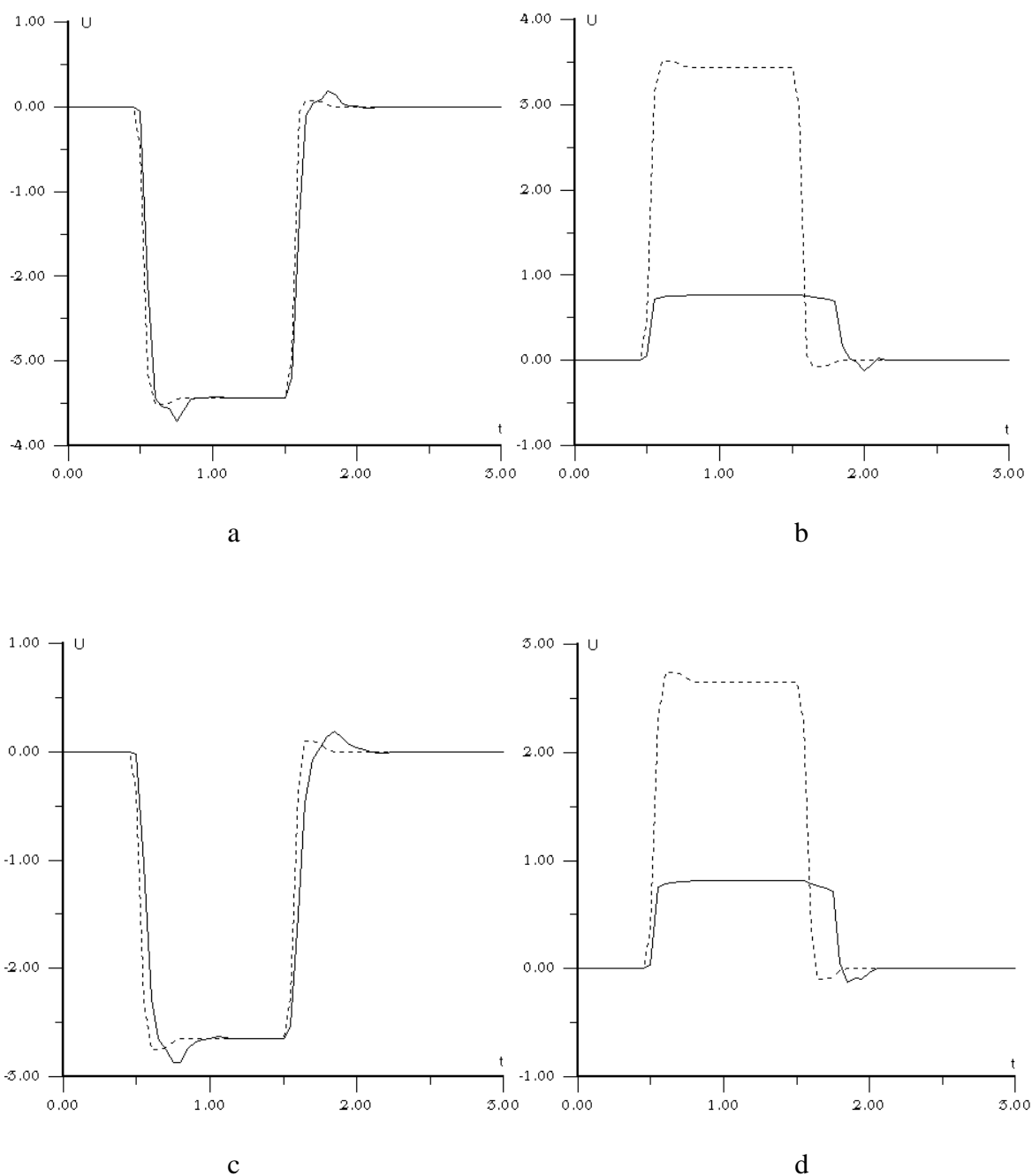

Fig. 7. Transients on the switch for multi-valued signals (U, V;  t, ps):

(a) Input signal- negative even mode pulse. The dotted line - signal at the input I, utter - on output II;

(b). Input signal -positive even mode pulse. The dotted line - signal at the input, utter - on output II;

(c) Input signal- negative odd mode pulse. The dotted line - signal at the input I, utter - on output III.

(d) Input signal- positive odd mode pulse. The dotted line - signal at the input I, utter - on output III.